\documentclass{aa}
\usepackage{amsmath,graphicx,amssymb}
\usepackage[varg]{txfonts}
\usepackage{natbib}
\usepackage{pst-eps}
\bibpunct{(}{)}{,}{a}{}{,}

\newcommand{\zav}[1]{\left(#1\right)}

\newlength\staretab

\newcommand\de{\text{d}}

\newcommand\hp{\ensuremath{H_\text{p}}}

\newcommand\msr{\ensuremath{M_\odot\,\text{yr}^{-1}}}

\newcommand\hvezda{\object{HD~191612}}
\newcommand\intvidpo{\!\!\int\limits_{\begin{array}{c}\text{\scriptsize
visible}\\[-2mm]\text{\scriptsize surface}\end{array}}\!\!}

\begin{document}

\title{The nature of the light variability of magnetic Of?p star HD 191612}

\author{J.~Krti\v{c}ka}


\institute{\'Ustav teoretick\'e fyziky a astrofyziky, Masarykova univerzita,
           Kotl\'a\v rsk\' a 2, CZ-611\,37 Brno, Czech
           Republic}

\date{Received}

\abstract{A small fraction of hot OBA stars host global magnetic fields with
field strengths of the order of 0.1\,--\,10\,kG. This leads to the creation of
persistent surface structures (spots) in stars with sufficiently weak winds as a
result of the radiative diffusion. These spots become evident in spectroscopic
and photometric variability. This type of variability is not expected in stars
with strong winds, where the wind inhibits the radiative diffusion. Therefore, a
weak photometric variability of the magnetic Of?p star HD~191612 is attributed
to the light absorption in the circumstellar clouds.}
{We study the nature of the photometric variability of HD~191612. We assume that
the variability results from variable wind blanketing induced by surface
variations of the magnetic field tilt and modulated by stellar rotation.}
{We used our global kinetic equilibrium (NLTE) wind models with radiative force
determined from the radiative transfer equation in the comoving frame (CMF) to
predict the stellar emergent flux. Our models describe the stellar atmosphere in
a unified manner and account for the influence of the wind on the atmosphere.
The models are calculated for different wind mass-loss rates to mimic the effect
of magnetic field tilt on the emergent fluxes. We integrate the emergent fluxes
over the visible stellar surface for individual rotational phases, and calculate
the rotationally modulated light curve of HD 191612.}
{The wind blanketing that varies across surface of HD 191612  is able to
explain a part of the observed light variability in this star. The mechanism is
able to operate even at relatively low mass-loss rates. The remaining
variability is most likely caused by the flux absorption in circumstellar
clouds.}
{The variable wind blanketing is an additional source of the light variability in
massive stars. The presence of the rotational light variability may serve as a
proxy for the magnetic field.}


\titlerunning{The nature of the light variability of magnetic Of?p star HD
191612}

\authorrunning{J.~Krti\v{c}ka}
\maketitle

\section{Introduction}

A small fraction of hot OBA stars host global magnetic fields with surface
intensities of the order of 0.1\,--\,10\,kG \citep{romag,wademim}. This leads to
the creation of persistent surface structures (spots) in stars with low
luminosity (weak or absent winds) as a result of the radiative diffusion
\citep{vasam,vimiri}. These spots cause prominent spectroscopic variability,
which is modulated by the stellar rotation \citep[e.g.][]{lukor,kolun}. The
inhomogeneous surface distribution of elements and redistribution of the flux in
the surface spots leads to the photometric variability of chemically peculiar
stars \citep{myteta,myfidra}.

This type of variability is not expected in magnetic stars with strong winds,
where the wind is predicted to inhibit the radiative diffusion
\citep{vasam,vimiri}. Instead, the magnetic field interacts with the ionized
stellar wind, which moves along the field lines. The effect of the magnetic
field on the stellar wind is characterized by the ratio between magnetic field
energy density and kinetic energy density of the wind. This may be parameterized
by the wind magnetic confinement parameter introduced by \citet{udo}. The larger
the magnetic field energy density is, the stronger is the influence of the
magnetic field on the flow. If the magnetic field energy density dominates, then
the structure of the flow depends on the relation between the Kepler corotation
radius $R_\text{K}$, at which the centrifugal force balances the gravity, and
the Alfv\'en radius $R_\text{A}$ \citep[see][]{udorot,malykor}, where the flow
velocity is equal to the Alfv\'en speed. The magnetosphere displays dynamical
phenomena for slowly rotating stars with $R_\text{A}<R_\text{K}$, whereas
circumstellar clouds exist in stars hosting centrifugal magnetospheres with
$R_\text{A}>R_\text{K}$.

Therefore, observed weak photometric variability of the magnetic Of?p star
\hvezda\ was attributed to the light absorption in the circumstellar clouds
rather than to surface spots \citep{wahot}. However, we argue that photometric
spots may exist even on the surface of hot stars with magnetic fields and winds.
These are not caused by the abundance inhomogeneities, but by wind mass flux
that is modulated by the different tilt of the magnetic field across the stellar
surface.

As the wind is forced to flow along the magnetic field in the strong field
limit, only a projection of the radiative force along the field lines is able to
accelerate the wind. Therefore, the surface wind mass flux depends on the
magnetic field tilt with respect to the surface normal and therefore on the location
on the stellar surface \citep{owoudan}. The wind blocks part of the photospheric
flux affecting the emergent flux and photometric colours \citep{acko}. This
effect is called wind blanketing and has to be modelled by a global atmosphere
models that unify the description of the photosphere and wind
\citep{hilmi,grakor,fastpuls}. Such models take the
feedback effect of the stellar wind on the photosphere properly into the account.

To our knowledge, the wind blanketing has never been studied as a source of the
light variability in magnetic O stars. For this purpose we selected Of?p star
\hvezda. This star has a relatively strong stellar wind \citep{marcoun}. The
period of weak photometric variability of this star of 536~days
\citep{koeyer,nazdis} is consistent with the period of spectral variability
\citep{walborn}. The discovery of a strong magnetic field with a polar strength
of about $-1.5$~kG \citep{donate} led to the current picture of rotationally
modulated variability of this star, which aims to explain the optical and
ultraviolet (UV) line variability \citep{sunhd,marcoun}, longitudinal magnetic
field variability \citep{wahot}, and the variability of X-ray flux
\citep{nazud}.

The photometric variability of \hvezda\ was successfully modelled with the Monte
Carlo radiative transfer code, which used the density structure from the MHD
simulations \citep{wahot}. The simulations adopted a relatively high wind
mass-loss rate $1.6\times10^{-6}\,\msr$, which was derived by \citet{howmoc}
from H$\alpha$ spectroscopy. However, the mass-loss rate derived from the UV
diagnostics $1.3\times10^{-8}\,\msr$ \citep{marcoun} is more than two orders of
magnitude lower. This difference likely reflects the problem of discordant
mass-loss rate determinations in O stars that is attributed to clumping
\citep[e.g.][]{pulchuch,sund,clres1,clres2}. Anyway, it would likely be problematic to explain the observed light variability with such a low wind
mass-loss rate. Therefore, we study the wind blanketing as a potential
additional source of the light variability in \hvezda.

\section{Global wind models}

We used our spherically symmetric stationary wind code METUJE (Krti\v cka \&
Kub\'at, in preparation) for the calculation of the wind models of \hvezda. The
code is based on our previous models \citep{cmf1}, however it is calculated in a
global (unified) approach that integrates the description of the hydrostatic
atmosphere and supersonic wind. The model radiative transfer equation is solved
in the comoving frame (CMF) with opacities and emissivities calculated using
occupation numbers that are derived from the kinetic equilibrium (NLTE) equations. For
given stellar parameters, the model enables us to predict consistently the
radial wind structure (i.e. the radial dependence of density, velocity, and
temperature) from hydrodynamical equations and to derive the wind mass-loss rate
$\dot M$ and the stellar emergent flux that accounts for the wind blanketing.

The ionization and excitation state was calculated from the NLTE equations. Part
of the corresponding models of ions \citep[see][for a complete list]{nlteiii}
was adopted from TLUSTY model atmosphere input files \citep{ostar2003,bstar2006}
and part was prepared by us using Opacity and Iron Project data
\citep{topt,zel0} and data described by \citet{pahole}. The level populations
were used to calculate opacities and emissivities in the CMF radiative transfer
equation. The radiative transfer equation was solved with a method of
\citet{mikuh} modified by \citet{cmf1}. We used the diffusion approximation as
the inner boundary condition. The solution of the radiative transfer equation
was used to calculate the radiative rates in the NLTE equations, the radiative force, and the radiative cooling and heating terms. For the
determination of the temperature we used three different methods depending on
the location in the atmosphere and wind. In the deepest layers of the
atmosphere we used the differential form of the radiative equilibrium, while in
the upper layers of the atmosphere we applied the integral form of the radiative
equilibrium \citep[e.g.][]{kubii};  in the wind we used the thermal balance
on free electrons method \citep{kpp}. The hydrodynamical equations (the
continuity equation, equation of motion, and the energy equation) were solved
iteratively together with NLTE and radiative transfer equations to obtain the
radial dependence of level populations, density, radial velocity, and
temperature. The line data used for the line force calculation were extracted
from the VALD database (Piskunov et al. \citeyear{vald1}, Kupka et al.
\citeyear{vald2}) supplemented for lighter elements (with atomic number
$Z\leq20$) using the data available at the Kurucz
website\footnote{http://kurucz.harvard.edu}. The initial model for the
iterations was derived from the combination of the model atmosphere calculated
with TLUSTY \citep{ostar2003} in the inner parts of the global model and in the outer parts of the
wind model.

\begin{table}[t]
\caption{\hvezda\ parameters adopted in this study}
\label{hvezda}
\centering
\begin{tabular}{lc}
\hline
Effective temperature ${{T}_\mathrm{eff}}$ & ${36\,000}$\,K \\
Radius $R_*$ & $14.1\,R_\odot$ \\
Mass $M$ & $29.2\,M_\odot$ \\
Predicted mass-loss rate $\dot M_0$ &  $2.6\times10^{-7}\,\msr$ \\
\hline
\end{tabular}
\end{table}

For our study we adopted the \hvezda\ parameters given in Table~\ref{hvezda}.
The effective temperature taken from \citet{marcoun} was used to derive mass and
radius from scalings of \citet{okali}. The adopted radius agrees with that
derived by \citet{howmoc} and \citet{marcoun} and the mass agrees with the estimate
of \citet{howmoc}. The mass-loss rate was predicted by our METUJE models
neglecting the effect of the magnetic field. Our predicted value is intermediate between the values derived from the optical and UV analysis
\citep{howmoc,marcoun}. We assumed solar chemical composition \citep{asp09}.

Our wind modelling neglects the influence of the magnetic field. In reality the
wind flows along the magnetic field lines, which affects its dynamics as a
result of the tilt of the flow with respect to the radial direction and modified
divergence of the flow \citep{udo}. However, we are concerned with the influence
of magnetic field on the wind blanketing, which is governed mostly by the wind
mass-loss rate. To model this effect, we artificially scale the radiative force
in the wind. This yields a series of wind models and emergent fluxes
$F(\lambda,\dot M)$ parameterized by the wind mass-loss rate.

\begin{figure}[t]
\centering
\resizebox{\hsize}{!}{\includegraphics{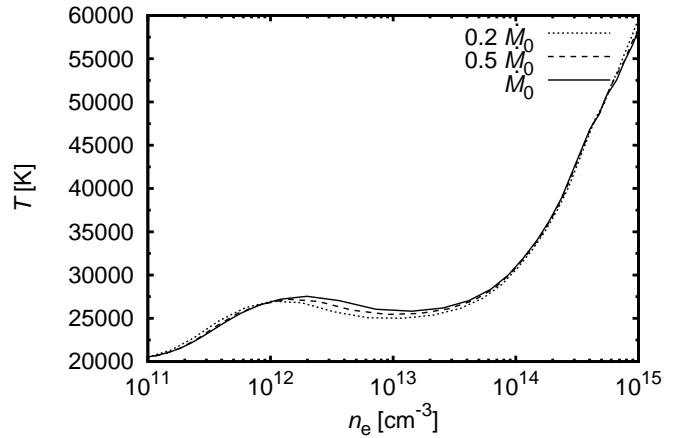}}
\caption{Dependence of the wind temperature as a function of the electron
density for models with different mass-loss rates.}
\label{tep}
\end{figure}

\begin{figure}[t]
\centering
\resizebox{\hsize}{!}{\includegraphics{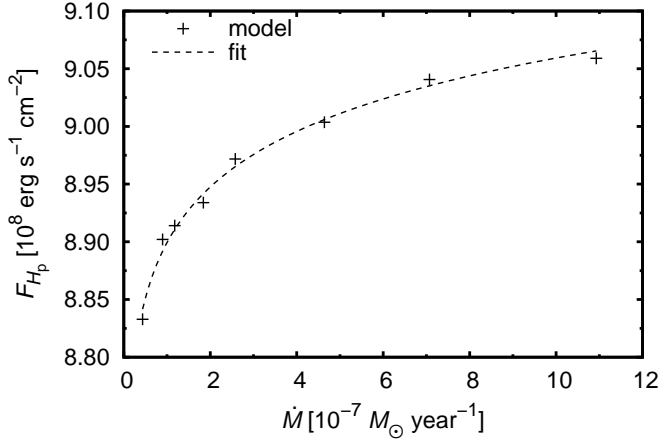}}
\caption{Dependence of the emergent flux in the \hp\ colour of the Hipparcos
photometric system on the wind mass-loss rate. Crosses denote the values from
the models and the dashed line fit Eq.~\eqref{stjacques}.}
\label{tok}
\end{figure}

The wind blanketing increases with increasing wind mass-loss rate. This results
in the increase of the temperature of the continuum forming region of the model
atmosphere (see Fig.~\ref{tep}) and causes the modification of the emergent flux
in the individual photometric passbands. Therefore, the \hp\ flux of the
Hipparcos photometric system $F\!_{\hp}(\dot M)$ depends on the wind mass-loss
rate, as shown in Fig.~\ref{tok}. The emergent flux in passband $\hp$ for a
given mass-loss rate was calculated as
\begin{equation}
F\!_{\hp}(\dot M)=\int_0^{\infty}\Phi_{\hp}(\lambda) \, F(\lambda,\dot M)
\, \text{d}\lambda,
\end{equation}
where $\Phi_{\hp}(\lambda)$ is the response function  of the
Hipparcos photometric system.
Fitting the emergent fluxes for different mass-loss rates we arrive at
approximate formula
\begin{equation}
\label{stjacques}
\frac{F\!_{\hp}(\dot M(\Omega))}{1\,\text{erg}\,\text{s}^{-1}\,\text{cm}^{-2}}=
1.6\times10^7 \log\zav{\frac{\dot M(\Omega)}{1\,\msr}}+1.0\times10^9,
\end{equation}
which gives an emergent flux in passband $\hp$ as a function of mass-loss rate.
Because we fitted the flux variations with a power law, the resulting
light curve is expected to be independent of an exact value of the mass-loss
rate. This shows that the studied mechanism can cause the light variability even
at a relatively low mass-loss rates.

\section{Modelling of the light variability of \hvezda}

We modified our code for the prediction of the light variability of chemically
peculiar stars \citep{myteta} to model the light variability due to the
wind blanketing. The magnitude difference between the flux $f_{\hp}$ at a given
phase and the reference flux $f_{\hp}^\text{ref}$ in passband $\hp$ is defined
as
\begin{equation}
\label{velik}
\Delta \hp=-2.5\,\log\,\zav{\frac{{f_{\hp}}}{f_{\hp}^\mathrm{ref}}}.
\end{equation}
The reference flux is obtained under the condition that the mean magnitude
difference over the rotational period is zero. The  radiative flux in a
passband $\hp$ at the distance $D$ from the spherical star is
\citep{hubenymihalas}
\begin{equation}
\label{vyptok}
f_{\hp}=\zav{\frac{R_*}{D}}^2\intvidpo I_{\hp}(\theta,\Omega)\cos\theta\,\text{d}\Omega.
\end{equation}
The intensity $I_{\hp}(\theta,\Omega)$ at angle
$\theta$ with respect to the normal at the surface was obtained at each surface
point with spherical coordinates $\Omega$ from the 
emergent flux taking into account the limb darkening $u_{\hp}(\theta)$
\begin{equation}
I_{\hp}(\theta,\Omega)=u_{\hp}(\theta)\,I_{\hp}(\theta=0,\dot
M(\Omega))=\frac{u_{\hp}(\theta)}{\langle u_{\hp} \rangle}
F\!_{\hp}(\dot M(\Omega))
,\end{equation}
with limb darkening coefficients from \citet{okraaj}. Here $\langle u_{\hp}
\rangle=2\pi\int_0^{\pi/2}u_{\hp}(\theta)\cos\theta \sin\theta\,\de\theta$.
The emergent flux
$F\!_{\hp}(\dot M(\Omega))$ as a function of mass-loss rate is taken from the
fit Eq.~\eqref{stjacques}. The
mass-loss rate was assumed to depend on the tilt of the magnetic field as
\citep{owoudan}
\begin{equation}
\dot M(\Omega)=\dot M_0 \cos^2\theta_\text{B}.
\end{equation}
This scaling was derived from analytical considerations, but it agrees with
numerical simulations.
The mass-loss rate $\dot M_0$ , which neglects the influence of the magnetic field
is given in Table~\ref{hvezda}.

For a dipolar magnetic field
\begin{equation}
\boldsymbol{B}=\frac{B_\text{p}R_*^3}{2}
\zav{\frac{3\boldsymbol{r}\zav{\boldsymbol{m}\cdot\boldsymbol{r}}}{r^5}-
    {\frac{\boldsymbol{m}}{r^3}}},
\end{equation}
where $B_\text{p}$ is the polar magnetic field strength, $\boldsymbol{r}$ is the
radius vector, and $\boldsymbol{m}$ is the unit vector in the direction of the
magnetic pole, the tilt of the magnetic field is given by a scalar product
\begin{equation}
\cos\theta_\text{B}=\frac{\boldsymbol{r}\cdot\boldsymbol{B}}{rB}=
2\zav{3\zav{\boldsymbol{m}\cdot\boldsymbol{r}}^2+r^2}^{-1/2}
{\boldsymbol{m}\cdot\boldsymbol{r}}.
\end{equation}
In the models, the tilt of the magnetic field $\theta_\text{B}$ was derived
assuming the dipolar field inclined by angle $\beta$ with respect to the
rotational axis. The axis of rotation is inclined by angle $i$ with respect to
the line of sight.

There is no unique measurement of the rotational axis inclination $i$ and
magnetic field obliquity $\beta$, however, the observations yield the sum of
these two values $i+\beta=95^\circ$ \citep{wahot}. Taking advantage of this
liberty, we selected parameters that fit the observed data best, i.e.
$i=50^\circ$ and $\beta=45^\circ$. The light curves with other parameters have
similar shapes but lower amplitudes.

\begin{figure}[t]
\centering
\resizebox{\hsize}{!}{\includegraphics{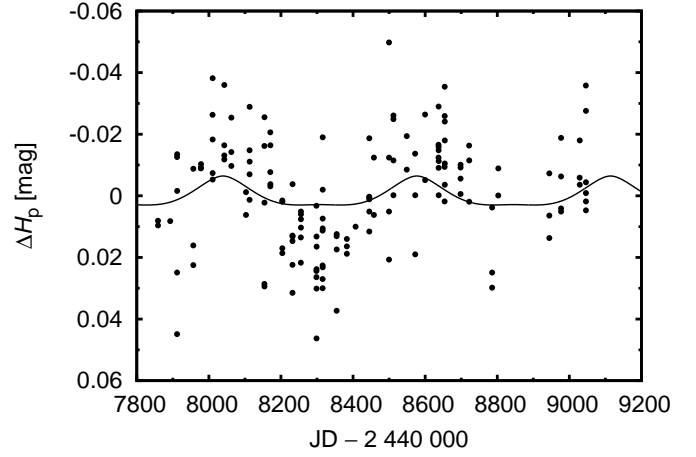}}
\caption{Comparison of simulated (solid line) and observed (Hipparcos, dots)
light curve of \hvezda. The light curve was plotted using the ephemeris of
\citet{howmoc}.}
\label{promhv}
\end{figure}

The observed and simulated light curves show the light maximum during the phases
when the magnetic pole is visible on the stellar surface (Fig.~\ref{promhv}).
The stellar wind flows freely along the magnetic field lines during these
phases, and consequently the mass-loss rate and the visual magnitude are the
highest. On the other hand, during the light minimum the magnetic equator appears
on the visible surface, and consequently the mass-loss rate is inhibited and the
visual flux is the lowest.

The observed and predicted light curves agree in shape. However, the observed
light curve has
amplitude $35\pm4$\,mmag, which is about three times
higher than the amplitude of the predicted light curve.
Consequently, our model is able to explain at most part of the light variability
of \hvezda.

\section{Conclusions and discussion}

We modelled the light variability of \hvezda\ assuming that the light
variability is caused by the influence of magnetic field on the wind mass flux.
The tilt of the magnetic field varies across the \hvezda\ surface, and consequently
the wind mass flux depends on the location on the stellar surface. As a result
of the wind blanketing, which increases with mass-loss rate, the emergent flux
is redistributed partly to the optical region. The stellar rotation modulates
the observed flux and leads to light variability.

The observed and predicted light curves agree in shape, but the observed light
curve has about three times higher amplitude than the predicted light
curve. The missing light variability can be explained by light absorption in
the circumstellar clouds \citep{wahot}.

The proposed mechanism is able to operate even at moderate mass-loss rates,
which are significantly lower than that inferred from the H$\alpha$ analysis.
The importance of the wind blanketing for light variability can be
observationally tested in the UV domain. Our calculations showed that
 light variability in the UV (e.g. around 1300\,\AA) is in the
anti-phase with visual variability.

The influence of the Zeeman effect on the opacity may also contribute to 
light variability in magnetic stars. However, the detailed model atmosphere
calculations with anomalous Zeeman effect and polarized radiative transfer
\citep{zeeman-paper2} performed for A stars showed that such an effect is important
only in stars with much stronger fields. The influence of this effect in \hvezda\
is therefore likely to be negligible.

The atmosphere of magnetic stars has a 3D structure and cannot be modelled
with spherically symmetric models in general. However, the most important effect
due to the magnetic field, i.e. the influence on the wind mass-loss rate, was
included in the models. The emergent flux is not significantly influenced by
the 3D effects, because the continuum forming region has significantly
smaller thickness than a typical scale at which the magnetic field varies. 

The study of the magnetic fields is especially important in O stars. Similar to
the chemically peculiar stars, light variability may serve as a proxy for
the search of the magnetic field using spectropolarimetry. The origin of the
magnetic field is still a matter of the debate. The magnetic fields dominates
the atmospheres of stars with a magnetic field of the order of 100~G and stronger,
but the magnetic field energy density is lower than the gas thermal energy
density inside the stars. Consequently, it is likely that the interstellar
magnetic field does not survive the initial violent period of star formation and
the surface magnetic field is most likely the descendant of the dynamo magnetic
field from the period of star formation \citep{bra}.

Magnetic A stars do not have strong winds, and consequently in these stars we
always observe the same layers during the main-sequence lifetime.
However, in O stars the material from the stellar atmosphere is blown away by
strong winds within days, and hence we always observe freshly exposed layers
of the stars. The detection of the magnetic field therefore shows that the
magnetic fields threads the deep subsurface layers of these stars. Moreover, the
study of time variability of this field (even just from the photometric
variations) can provide information about the magnetic field variation inside
the stars.

\begin{acknowledgements}
This research was supported by grant GA\,\v{C}R  16-01116S.
Access to computing and storage facilities owned by parties and projects
contributing to the National Grid Infrastructure MetaCentrum, provided under the
programme Projects of Large Infrastructure for Research, Development, and
Innovations (LM2010005), is greatly appreciated.
\end{acknowledgements}

\end{document}